\documentclass[journal]{IEEEtran}

\ifCLASSINFOpdf
\else
   \usepackage[dvips]{graphicx}
\fi
\usepackage{url}

\usepackage{graphicx}
\usepackage[namelimits]{amsmath} %数学公式
\usepackage{amssymb}             %数学公式
\usepackage{amsfonts}            %数学字体
\usepackage{mathrsfs}            %数学花体
\usepackage{bm}            %加粗
\usepackage{subfigure}
\usepackage{caption}
\usepackage{multirow}
\usepackage{bbding}
\usepackage{cite}
\usepackage{hyperref}

\begin{document}

\title{Edge Heuristic GAN for Non-uniform Blind Deblurring}

\author{Shuai Zheng, Zhenfeng Zhu, Jian Cheng, Yandong Guo, and Yao Zhao, \IEEEmembership{Senior Member, IEEE}
\thanks{This paragraph of the first footnote will contain the date on which you submitted your paper for review. It will also contain support information, including sponsor and financial support acknowledgment. For example, ``This work was supported in part by the U.S. Department of Commerce under Grant BS123456.'' }
\thanks{S. Zheng, Z. Zhu, and Y. Zhao are with the Institute of Information Science, Beijing Jiaotong University, Beijing 100044, China, and also with the Beijing Key
	Laboratory of Advanced Information Science and Network Technology,
	Beijing 100044, China (e-mail: zs1997@bjtu.edu.cn; zhfzhu@bjtu.edu.cn;
	yzhao@bjtu.edu.cn).}
\thanks{J. Cheng is with the National Laboratory of Pattern Recognition, Institute of Automation, Chinese Academy of Sciences and University of Chinese
	Academy of Sciences, Beijing 100190, China (e-mail: jcheng@nlpr.ia.ac.cn).}
\thanks{Y. Guo is with the Xpeng Motors, Beijing 100080, China (e-mail: guoyd@xiaopeng.com).}}
\markboth{Journal of \LaTeX\ Class Files, Vol. 14, No. 8, August 2015}
{Shell \MakeLowercase{\textit{et al.}}: Bare Demo of IEEEtran.cls for IEEE Journals}
\maketitle

\begin{abstract}

Non-uniform blur, mainly caused by camera shake and motions of multiple objects, is one of the most common causes of image quality degradation. However, the traditional blind deblurring methods based on blur kernel estimation do not perform well on complicated non-uniform motion blurs. However, recent studies show that GAN-based approaches achieve impressive performance on deblurring tasks. In this letter, to further improve the performance of GAN-based methods on deblurring tasks, we propose an edge heuristic multi-scale generative adversarial network(GAN), which uses the “coarse-to-fine” scheme to restore clear images in an end-to-end manner. In particular, an edge-enhanced network is designed to generate sharp edges as auxiliary information to guide the deblurring process. Furthermore, We propose a hierarchical content loss function for deblurring tasks. Extensive experiments on different datasets show that our method achieves state-of-the-art performance in dynamic scene deblurring.
\end{abstract}


\IEEEpeerreviewmaketitle

\section{Introduction}

\IEEEPARstart{M}{otion} blur is one of the most common problems in computer vision. Camera shake and fast objects motions will cause non-uniform blurring, resulting in degraded image quality. How to effectively estimate unknown sharp image from given blurred image has always been one of the hotspots of researchers. Early studies mostly dealt with simple blurring caused by camera motions such as translation or rotation. Recent works have focused on handling non-uniform dynamic blurs caused by object motions and camera shakes.

Most of the deblurring methods use the physical model to generate clear images by estimating the blur kernel and latent images. However, it is worth noting that finding a blur kernel for each pixel is an ill-posed problem. Numerous methods \cite{1,2,3,4,5,6} use different statistical priors to help estimate blur kernels and latent images. For example, Xu et al. \cite{6} proposed an L0 sparse expression as a prior to jointly estimate sharp image and blur kernels.The dark channel is also used as a prior to capture intermediate latent image to aid in blur kernel estimation\cite{9}. Some recent work has applied CNN to image blind deblurring\cite{10, 11, 12, 13, 14, 15}. These methods all perform kernel estimation steps through CNN for restoring the latent sharp image. Sun et al. \cite{11} proposed an energy model consisting of CNN probability prior and smoothing prior, to estimate a smooth variable blur kernel. Extensive experiments show that both optimization-based and learning-based approaches have poor performance when dealing with non-uniform blurred images, and the operation speed is slow, which cannot meet the real-time requirements. 

In recent years, generative adversarial network(GAN)\cite{2014Generative} has achieved excellent performance in a variety of computer vision tasks\cite{18,19,20}.
GAN is also applied to image blind deblurring tasks due to the advantages in fitting data distribution\cite{16,17,23}. A kernel-free method proposed by Nah et al.\cite{16} uses the GAN framework to directly blur images and has good performance. Ramakrishnan et al. \cite{23} use the combination of pix2pix framework \cite{21} and densely connected convolutional networks \cite{22} to restore latent sharp images from blurred images. In addition, Kupyn et al.\cite{17}  proposed an end-to-end learned method for motion deblurring which is based on a conditional GAN and perceptual loss\cite{2016Perceptual}. These GAN-based methods have achieved good results in non-uniform blurred scenes, but there are still some problems to be solved, such as image details being over-smoothed, and artifacts still exist when dealing with violent blurring.

In this letter, we propose an edge heuristic multi-scale GAN to achieve non-uniform blind deblurring. The proposed network consists of two stages and does not involve blur kernel estimation. In the first stage, a lightweight conditional GAN called Edge-generated Net receives blurred edge and generates corresponding sharp edge. Then, the sharp edge is input as auxiliary information together with the corresponding blurred image into a multi-scale network. In addition, we have designed a new combined content loss to ensure that the network can achieve the desired results. Extensive experiments on challenging datasets show that our method has good performance in dynamic scene deblurring. 
%本文中，我们提出了一种阶段式多尺度生成对抗网络，其中，第一阶段使用encoder-decoder网路去尽可能生成清晰地边缘，并将其与模糊图像一起输入到第二阶段的多尺度网络中来约束和引导生成清晰图像。此外，我们融合暗通道先验、感知损失以及L2损失为multi-component loss function，这保证了高层特征的内容不变性以及确保可以生成更好的纹理细节。在具有挑战性的数据集上的实验表明，我们的网络具有良好的性能。
\vspace{-2mm}
\section{approach}
\begin{figure*}[ht]
	
	\centering
	\includegraphics[width=5.5in]{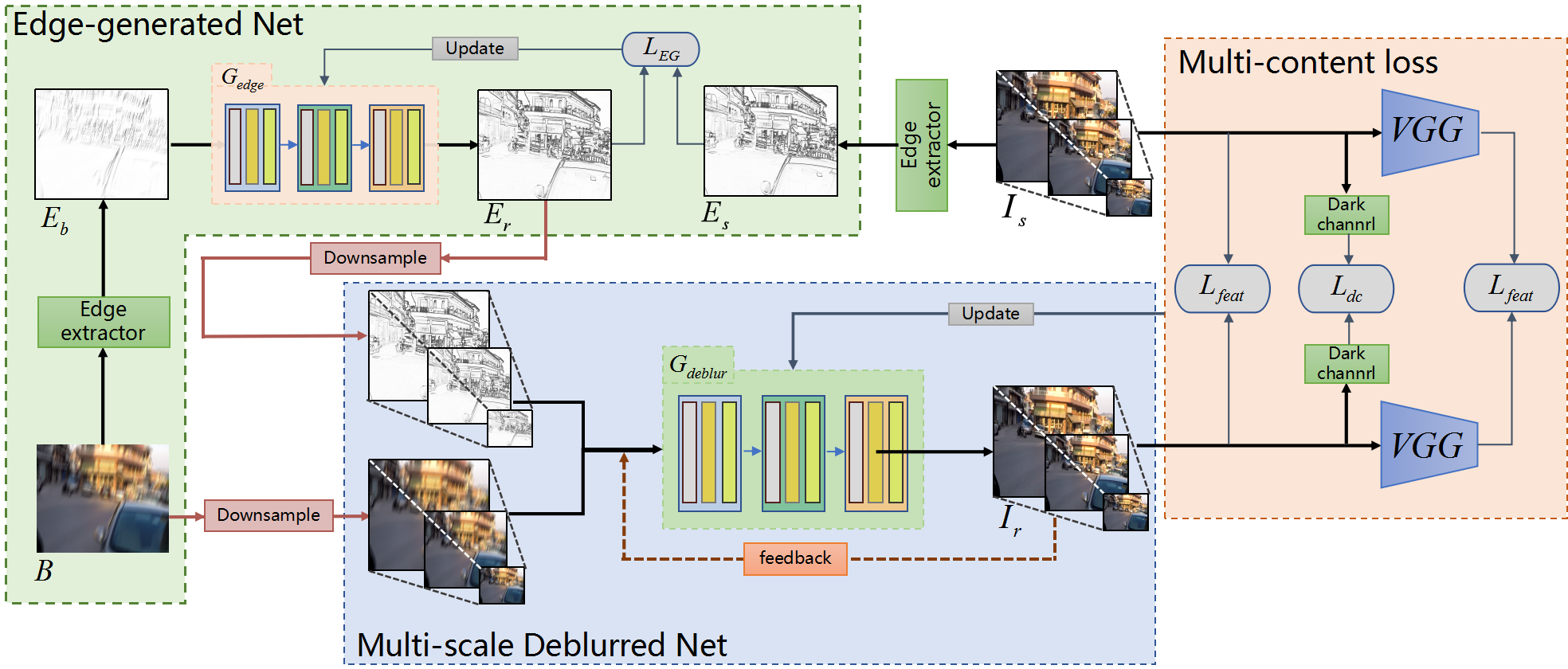}
	\caption{Overview of our architecture. $E_{b}$, $E_{r}$, $B$, $I_{r}$, and $I_{s}$ denote blurred edges, restored edges, blurred images, restored images, and clear images, respectively. Edge-generated Net receives $B$ and generates corresponding $E_{r}$. Multi-scale Deblurred Net uses $B$, coarse image and $E_{r}$ to generate refined image. The red dashed line indicates that the low-scale image is upsampled as part of the input for the next scale. }
	\label{fig1}
	
\end{figure*}

The overall framework of the proposed GAN is illustrated in Fig.\ref{fig1}.  The network uses the "coarse-to-fine" scheme to generate refined images with coarse images and generated edges as auxiliary information.
\vspace{-2mm}

\subsection{Edge-generated Net}
The blurring process weakens the high-frequency part of images to a certain extent, which mainly represents the edge of images. Besides, the widely used multi-scale network\cite{16} is also proved to weaken edge information during downsampling. However, strong edges play an important role in subjective image quality evaluation. Just as heuristic edge selection steps are often required in the MAP framework, the sharp image edge can be used as strong supervised information to assist image deblurring while preventing high-frequency information from being over-smoothed. So we carefully design a lightweight GAN to generate sharp edges which we call heuristic edges. 

Due to the excellent performance of the encoder-decoder network architecture in a variety of computer vision tasks\cite{2016Deep,2017Deep,2014Coarse}, we adopt a similar network architecture as a generator, but with some modifications to meet the task requirements. To ensure that the network can handle complicated motion blur, we add ResBlocks\cite{he2016deep} behind each convolutional layer and in front of the deconvolution layer to enhance the learning ability of the network. Each ResBlock contains two convolution operations, which provide a larger receptive field and prevent the vanishing gradient problem. For the deblurring task, the generated result mainly depends on the image instance currently input, therefore,  the BN layer is not suitable for this task. Instead, the BN layer in the ResBlock is replaced by the instance normalization(IN)\cite{2016Instance} layer, which not only speeds up model convergence but also maintains independence between each image instance. In addition, like U-Net\cite{2015U}, we retain the skip connection between the corresponding feature maps, which benefits more information about the original image texture to propagate in high-resolution layers. 
The discriminator architecture is the same as that used in deblurGAN. The edge extracted from the blurred image acts as generator input, and the output is corresponding sharp edge.
%The detailed parameters of the Edge-generated Net are as follows. 1 input block and 2 convolution blocks constitute the encoder, 2 deconvolution blocks and 1 output block constitute the decoder. Between encoder and decoder is a single ResBlock. Input block produces 32-channel feature map. The number of kernels of the two convolution blocks is 64,128 respectively. For deconvolution blocks, they are 128 and 64. In order to ensure the validity of the input and output content, the normalization layer is not adopted in input block and output block.
%\setlength{\belowcaptionskip}{-0.5cm} 
\begin{figure*}[htb]
	\centering
	\setlength{\abovecaptionskip}{-0.2cm}  
	\setlength{\belowcaptionskip}{-0.5cm}
	\subfigure[Blurry Image]{
		\centering
		\includegraphics[width=2in]{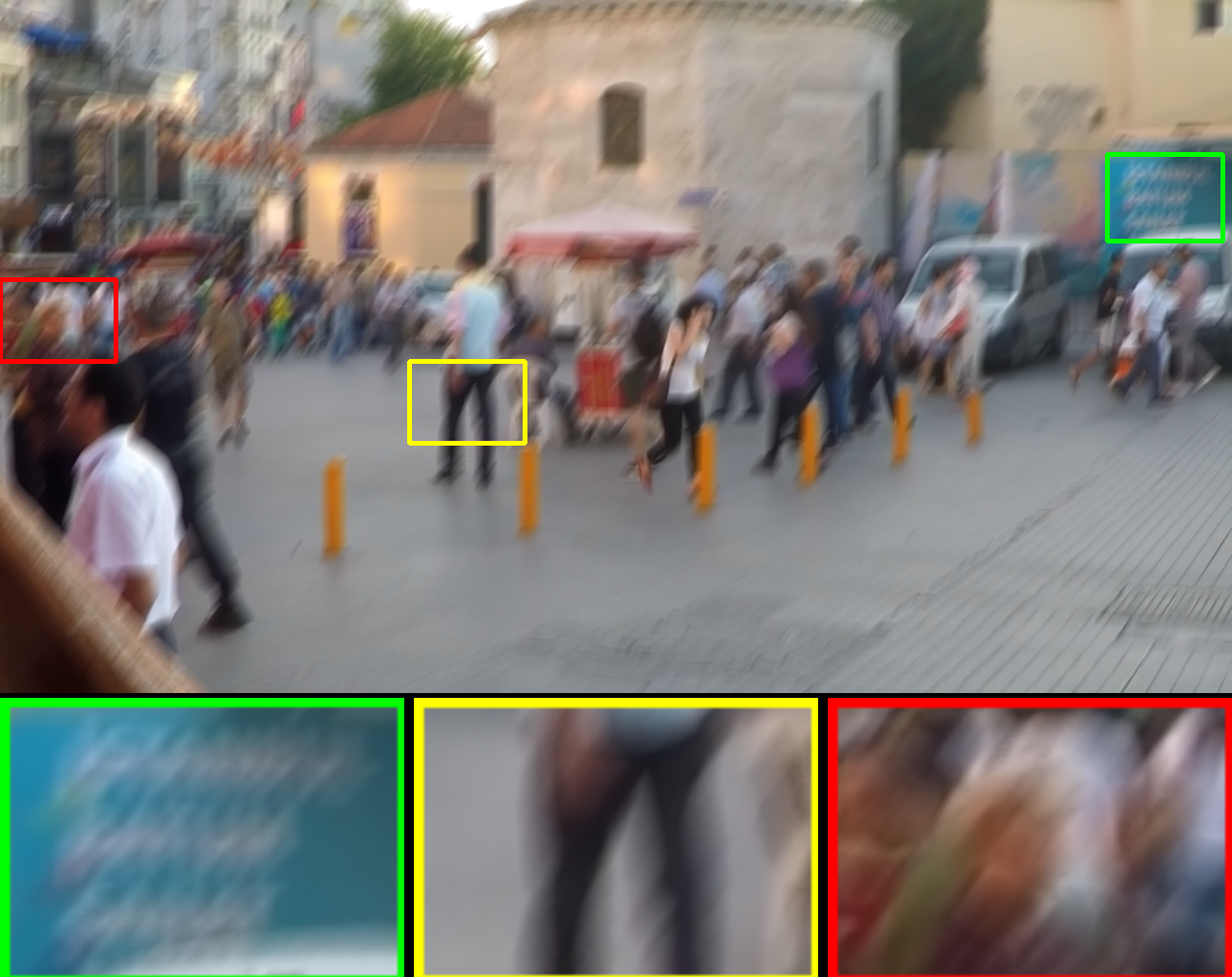}
		%\caption{fig1}
	}%
	\subfigure[Pan\cite{9}]{
		\centering
		\includegraphics[width=2in]{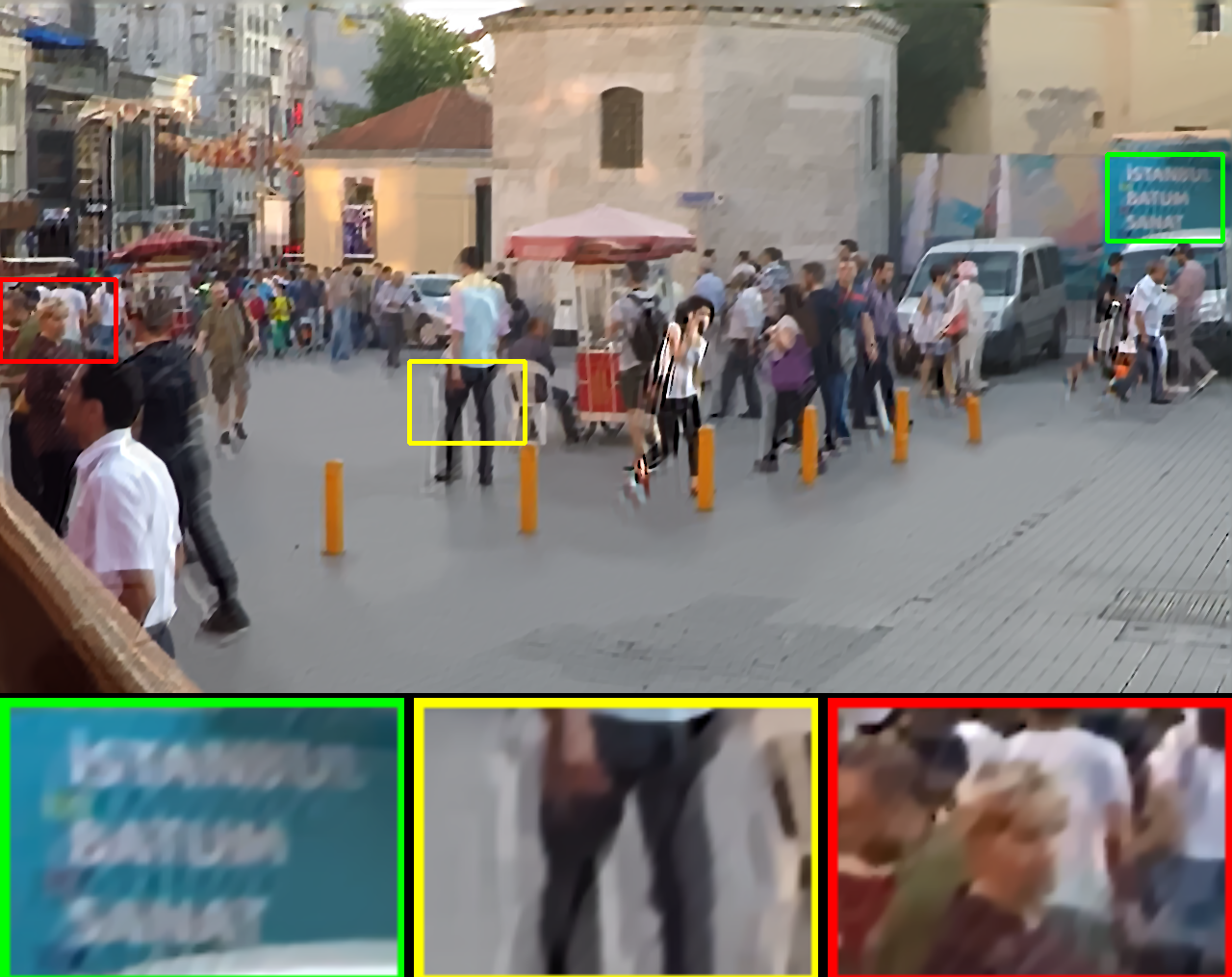}
		%\caption{fig2}
	}%
	\subfigure[Nah\cite{16}]{
		\centering
		\includegraphics[width=2in]{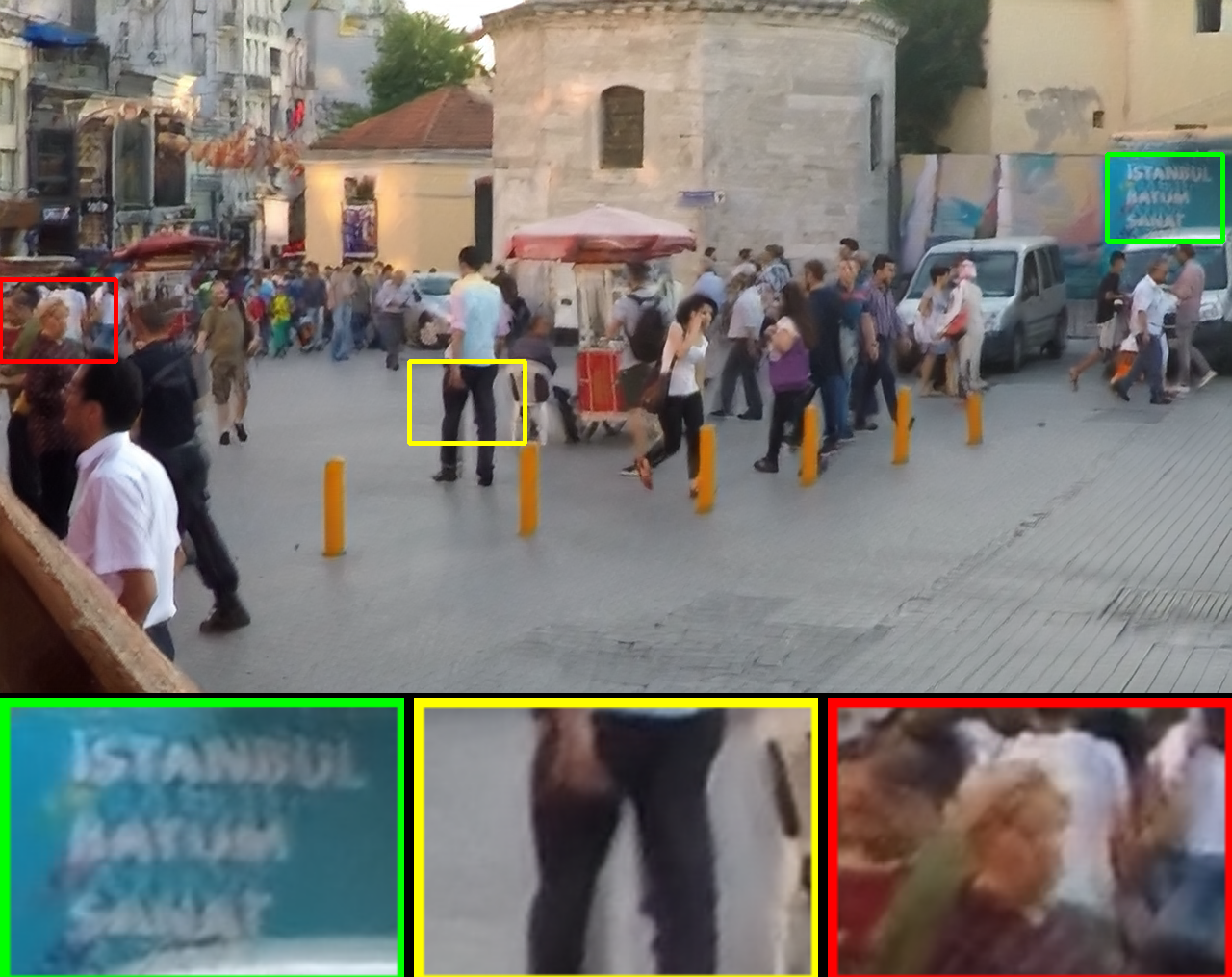}
		%\caption{fig2}
	}%
	
	\subfigure[Kupyn\cite{17}]{
		\centering
		\includegraphics[width=2in]{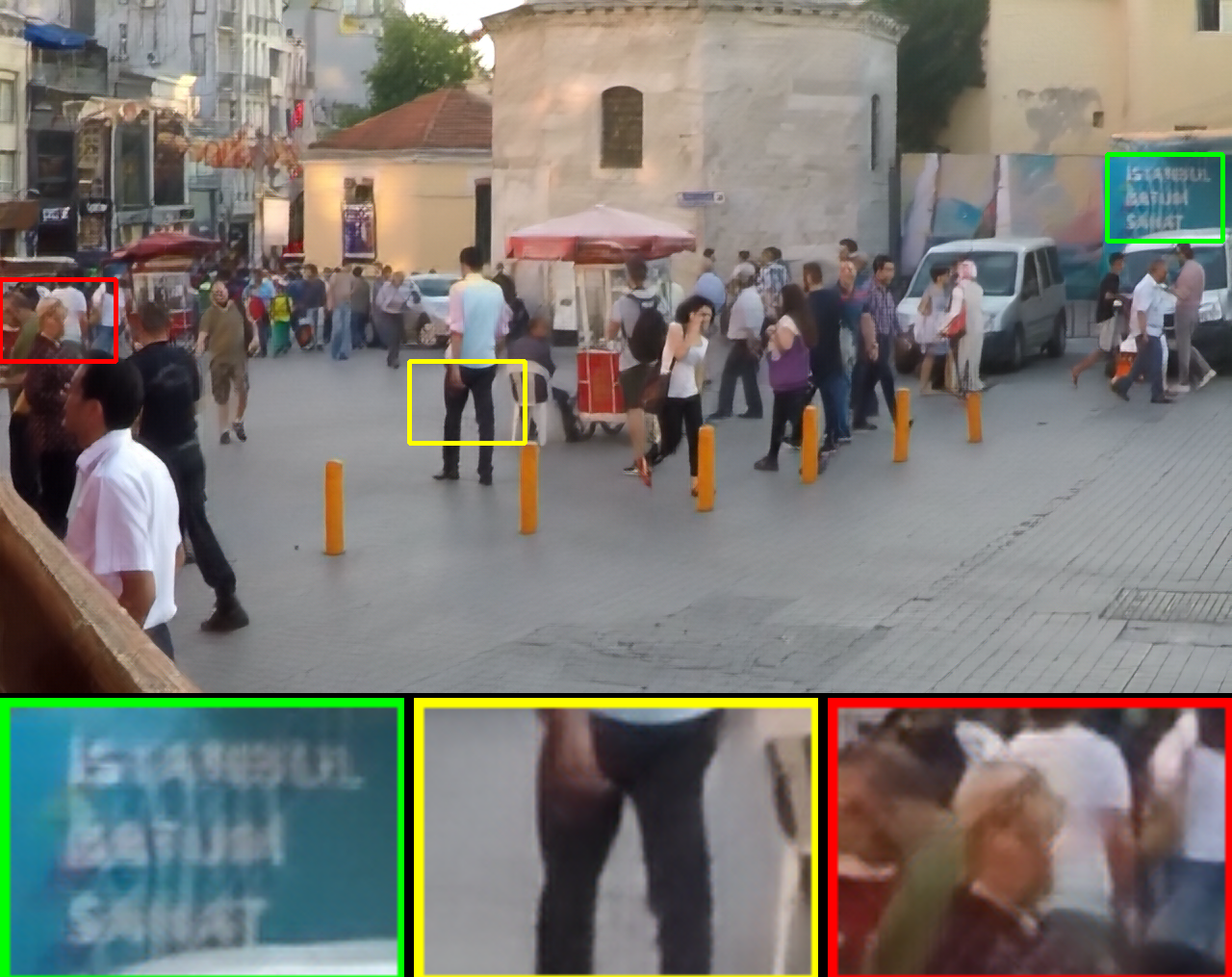}
		%\caption{fig1}
	}%
	\subfigure[Our]{
		\centering
		\includegraphics[width=2in]{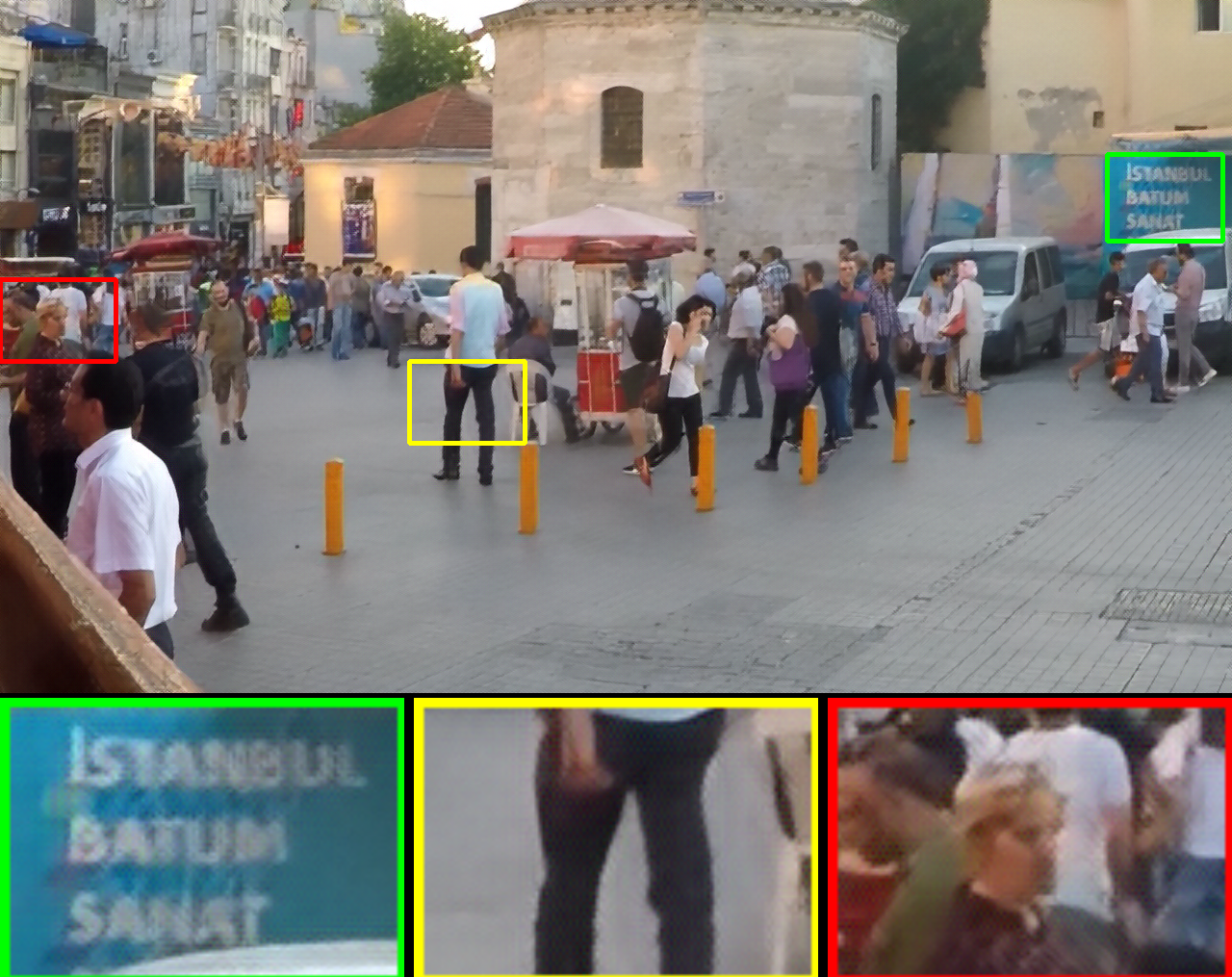}
		%\caption{fig2}
	}%
	\subfigure[Sharp Image]{
		\centering
		\includegraphics[width=2in]{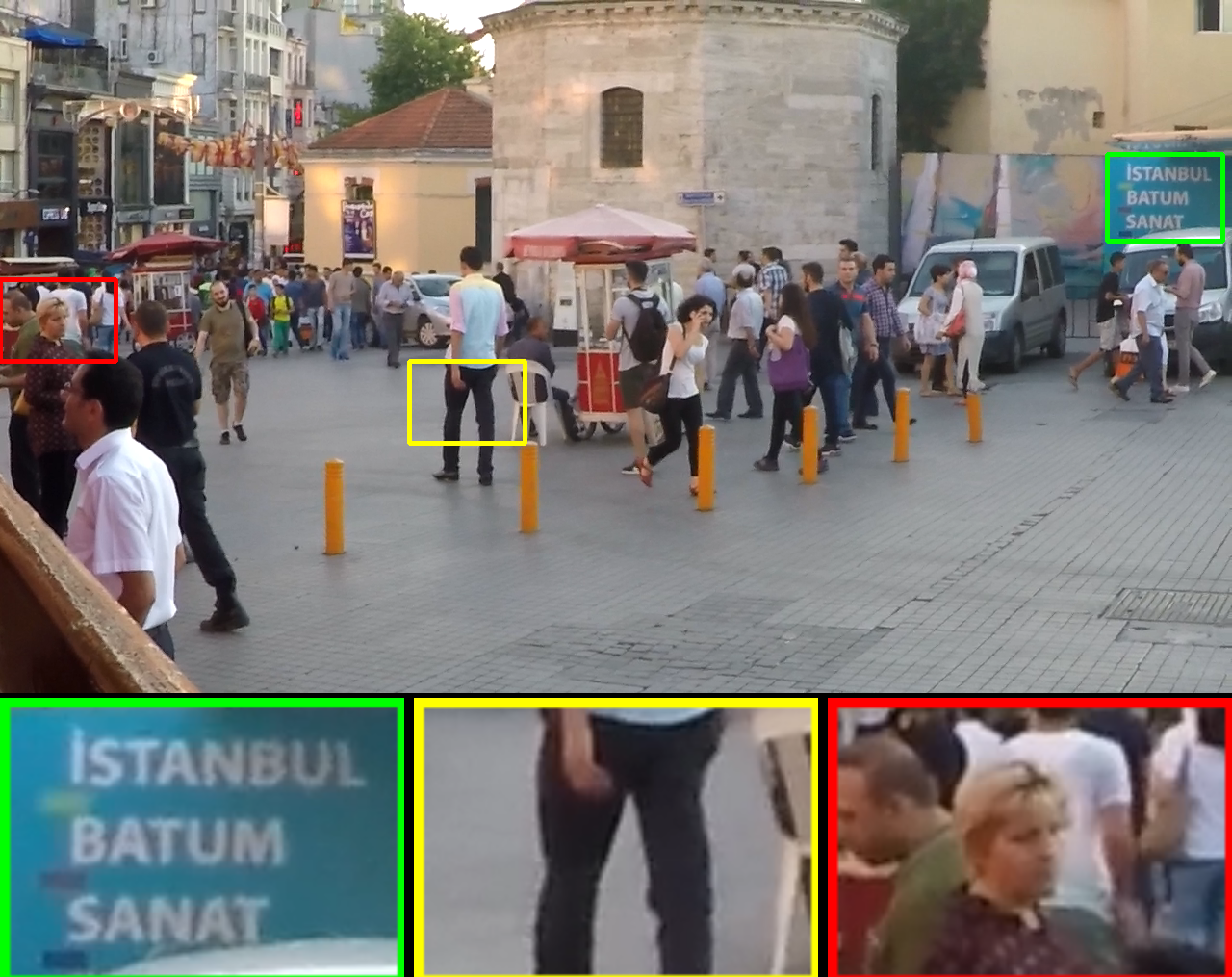}
		%\caption{fig2}
	}%
	
	\centering
	\caption{\textbf{Visual comparisons on GoPro testing dataset.} We show our deblurring result compared with the state-of-the-art methods\cite{9,16,17}.Our method not only effectively eliminated artifacts, but also restored good details and edges.}.
	\label{fig2}
\end{figure*}

\subsection{Multi-scale Deblurred Net}
Multi-scale network architecture has proved its effectiveness in many image enhancement tasks\cite{16}\cite{2017High}. In particular, it has been observed that a sufficiently low-resolution downsampled image of a blurred image approximates a sharp image of the corresponding resolution. Here, we design a multi-scale network as GAN generator for image deblurring. The input of each scale of the multi-scale generator is the downsampled blurred image, corresponding sharp edge and the output of the coarser scale. So the output of the generator can be expressed as 
\begin{equation}
\setlength{\abovedisplayskip}{3pt}
\setlength{\belowdisplayskip}{3pt}
I^{i} = G_{deblur}(B^{i}, I^{i-1}, E; \theta_{G} )
\end{equation}
where $ B^{i} $, $ I^{i} $, and $ E $ denote the blurred image , latent sharp image, and edge image at the i-th scale, respectively. $ G_{deblur} $ is the generator with training parameters denoted as $\theta_{G}$. Since each scale generator is designed to generate a sharp image, the discriminator is parameter-shared for generators of different scales, which can effectively improve the training speed.

Here are details of the network. We use convolution/deconvolution-IN-ReLu modules as ConvBlock and DeconvBlock\cite{2015U}. An InBlock transforms the input image into 64-channel feature map and two ConvBlocks stacked followed by input block that transforms the feature map into 256-channel feature map. Then, the feature map is input into 6 Resblocks to increase the receptive field. Finally, 2 DeconvBlock and 1 OutBlock transforms the feature map to the original input size. The kernel size of convolutional layer in ConvBlock is 5, and the stride size is 2. However, in order to overcome the checkerboard effect and improve the image quality, similar to sub-pixel convolution\cite{shi2016real}, the kernel size of deconvolutional layer in DeconvBlock is 4, and the stride size is still 2. The generator structure of each scale is the same
\vspace{-3mm}
%图像去模糊网络中，每个层级的输入是由下采样的模糊图像与相应的回复过的边缘，concat with 上一层的输出。我们期望的输出是网络可以表示为：公式X，解释公式。由于每一个level的network都是为了生成对应尺度下的清晰图像，所以每个层级共享同一个鉴别器，这样可以有效提高训练速度。
%生成器的每个层级架构都与deblurGAN相似，介绍网络具体参数。
%但是为了克服棋盘效应，提高图像质量，deconv层中的卷积核大小被设置4，步长为2
\subsection{Loss}
We train our GANs through two loss functions: content loss and adversarial loss.
%我们通过两个损失函数来训练我们的网络：内容损失和对抗损失。
\subsubsection{Adversarial loss}
For GAN, discriminator and generator play the following two-player minimax game with value function V (G; D). Arjovsky et al. \cite{2017Wasserstein} discuss the difficulties in GAN training caused by JS divergence approximation and propose WGAN using Wasserstein-1 distance：
\begin{align}
\underset{G}{\textup{min}}\ \underset{D}{\textup{max}}V(D,G)=&\mathbb{E}_{x\sim P_{data}(x)}[D(x)]- 
\nonumber\\ &\mathbb{E}_{z\sim P_{z}(z)}[D(G(z))]
\end{align}
where $p_{data}$ and $p_{z}$ denote the real data distribution and the noise distribution, respectively.
And adding gradient penalty\cite{2017Improved} terms can enforce Lipschitz constraint in WGAN:
\begin{align}
\mathbb{E}_{\tilde{x}\sim P_{g}(\tilde{x})}[\left \| \bigtriangledown_{\tilde{x}}D(\tilde{x})  \right \|_{2}-1]
\end{align}
where $ P_{g}$ is the generated data distribution, defined by
$\tilde{x} = G(z)$, $ z \sim P_{z}(z)$.

 We use WGAN-GP as the critic function which converges quickly and has high hyperparameter sensitivity. And for the sake of description, we use $G$ to represent $G_{deblur}$. Hence, the adversarial loss for edge generation network is defined as:
\begin{align}
L_{adv1}=-D(G_{edge}(E_{b}))
\end{align}
where $E_{b} $ represents the edge extracted from the blurred image. And we use $E$ to represent the edge image generated by pre-trained $G_{edge}$, so the adversarial loss for the i-th scale of multi-scale network can be defined as:
\begin{align}
L_{adv2}=-D(G(B^{i}|I^{i-1},E))
\end{align}
%在实践中，V (G; D)通常用交叉熵来求解以指导D和G+
%的更新。Arjovsky et al. [2] discuss the difficulties in GAN training caused by JS divergence approximation and propose WGAN using Wasserstein-1 distance：
%公式WGAN。在真实样本和生成样本之间进行插值，增强WGAN中莱布尼兹约束，这可以由梯度惩罚实现。WGANloss使生成器无需过多超惨调整即可得到优异的性能表现。

\subsubsection{Hierarchical content loss}
%对于每个生成器而言，我们组合了MSEloss、perceptloss以及暗通道损失作为内容损失。整体公式。
Hierarchical content loss can be decomposed into three parts: general content loss, texture loss, and dark channel loss. First, we use MSE loss to ensure general content consistency. 
\begin{align}
L_{pixel}(y,\hat{y})=\frac{1}{CHW}\left \| y-\hat{y}) \right \|_{2}^{2}
\end{align}
where $y$ is the input image, $\hat{y}$ is the generated image, and their shape is $C\times H\times W$. However, the MSE loss smooths the image and is not sufficient to handle the artifacts caused by large-motion blur. Hence, we add  Perceptual loss\cite{2016Perceptual} to the loss function. Perceptual loss is the Euclidean distance between feature maps extracted by pre-trained CNN of target images and generated images:
\begin{align}
L_{feat}(y,\hat{y})=\frac{1}{C_{j}H_{j}W_{j}}\left \| \phi _{j}(y)-\phi _{j}(\hat{y})) \right \|_{2}^{2}
\end{align}
where $ \phi _{j}(y)$ is the feature map of shape $C_{j}\times H_{j}\times W_{j}$ for the input $y$, which is obtained by the j-th convolution block within the pretrained VGG-19 network. We extracted the feature map of the conv3\_3 layer so that the perceptual loss takes into account both general content and texture details.

The pixel value of the dark channel increases due to the motion blur averaging of dark pixels and adjacent pixels, hence, the sparse difference of the dark channel reflects the severity of the blur. Unlike the processing in\cite{9},  MAE loss is used to measure the difference of dark channel map between ground truth and deblurred image

%Pan found that the dark channel of the blurred image changes and introduces it into the deblur task to suppress artifacts. Unlike the treatment of pan,  MAE loss is used to measure the difference of dark channel map between ground truth and deblurred image:
\begin{equation}
L_{dc}(y,\hat{y})=\frac{1}{HW}\left \| \varphi (y)-\varphi(\hat{y})) \right \|_{1}
\end{equation}
Finally, the generator network and discriminator
network is jointly trained by combining the content loss and adversarial loss. for the $G_{edge}$, final loss term is
\begin{equation}
L_{EG}=L_{adv1} +\lambda L_{feat}(E_{s},E_{r})
\end{equation}
And for the $G_{deblur}$, final loss term is
\begin{align}
&L_{MS}=L_{adv2} +\lambda L_{cont}
\\ &L_{cont}=L_{feat}(I_{s}^{i},I_{g}^{i})+L_{pixel}(I_{s}^{i},I_{g}^{i})+ L_{dc}(I_{s}^{i},I_{g}^{i})
\end{align}
where $E_{s}$ and $E_{r}$ denote edge extracted from ground truth and generated edge image, respectively. The weight constant $\lambda = 100$.
\vspace{-3.5mm}
\section{experiment}
\label{sec:guidelines}
Our experiments are conducted on a PC with Intel Xeon E5 CPU and an NVIDIA GTX 1080Ti GPU and the network is implemented on the Tensorflow platform.
\vspace{-3.5mm} 
\subsection{Data Preparation}
The GoPro dataset was used for training and testing, and the Kohler benchmark was also used to assess the generalization capabilities of the model.

\subsubsection{GoPro dataset}Nah et al.\cite{16} created the GoPro dataset that generates blurred images by averaging successive short exposure frames to approximate long exposure blurred frames. It simulates complex camera shake and object motion, and is more suitable for blurry images in real scenes. Following the same strategy as in \cite{16}, we use 2,103 pairs for training and the remaining 1,111 pairs for evaluation.

\subsubsection{K\"ohler benchmark}K\"ohler benchmark\cite{Rolf2012Recording} is a standard
benchmark for evaluation of blind deblurring, consists of 4 latent images and corresponding 48 blurred images which blurs are caused by replaying recorded 6D camera motion. We also used the K\"ohler benchmark as a test set for model performance evaluation. 
% 研究人员[25, 33]提出从高速摄像机(如GoPro Hero 4 Black)拍摄的视频中，通过平均连续的短曝光帧来生成模糊图像，以近似长曝光的模糊帧。该数据集模拟了复杂的相机抖动和物体运动，更加贴合真实场景下的模糊图像。为了在网络结构上进行公平的比较，我们使用了GOPRO的[25]数据集来训练我们的网络，which contains 22 sequences with 2, 103 blurred/clear image pairs，。该数据集已经分好了训练集与测试集，其中训练集中有2103个图像对，测试集中有1111个图像对。除非特殊说明，我们的所有实验都是在具有相同PC配置的训练集进行训练的。
\vspace{-3.5mm}
\subsection{Training details}
We follow the training strategy in WGAN-GP, using Adam as the optimizer with β1 = 0.9, β2 = 0.999, and perform 5 gradient descent steps on $D$, then one step on $G$. The learning rate is set initially to 10−4 and decayed to 10-6 at 600 epochs for both generator and critic of Edge-generated Net and Multi-scale deblurred Net. Similar to various CGANs, all the models were trained with a batch size = 1. Since the models are fully convolutional, they can be applied to images of arbitrary size. Unless otherwise stated, all of our experiments were trained on GoPro training set with the same PC configuration. 
 
%我们遵循WGAN-GP中的训练策略，使用Adam作为solver with β1 = 0.9, β2 = 0.999，每进行五次鉴别器的更新，进行一次G的更新。The learning rate is set initially to 10−4 for both generator and critic。在600次epoch时衰减到XX。与大多数CGAN类似，All the models were trained with a batch size = 1。由于我们的网络是完全卷积的，因此任意大小的图像都可以作为输入。
	\begin{table}[htb]
	\centering
	\small 
	\caption{Quantitative results of the baseline models}
	\renewcommand{\arraystretch}{1.3}
	\begin{tabular}{l|l|l|l|l}
		\hline
		& \textbf{B} & \textbf{BE}    & \textbf{BC}    & \textbf{Proposed} \\ \hline
		Edge-generated Net    & $\times$        & \checkmark     & $\times$     & \checkmark        \\
		Combined content loss & $\times$        & $\times$     & \checkmark     & \checkmark        \\ \hline
		PSNR                  & 28.30    & 28.81 & 28.73 & 29.06    \\ \hline
	\end{tabular}
\end{table}
\vspace{-5mm}
\subsection{Ablation Study}
%表格deblurGAN，加边缘，加暗通道，加L2
Unlike conventional data expansion, high frequency information of the image and images of different scales are introduced into the proposed frame to assist in image deblurring. Here, we design three baseline models to verify the effectiveness of edge constraint and combined content loss. Table I shows the investigation on the effects of edge constraint and combined content loss. Model B that trained by adversarial loss and perceptual loss use the same architecture as our proposed one, but it doesn't have restored edge as auxiliary information. After adding the edge generation network to the model \textbf{B}, the PSNR performance of the model \textbf{BE} is improved to 28.81 and when we replaced the simple perceptual loss with the combined content loss, the PSNR performance of the model \textbf{BC} is improved to 28.73. We trained the four networks with the same training configuration and hyperparameter, which shows that these two parts can effectively improve model performance.

\begin{figure}[htb]	
	\centering
	\setlength{\belowcaptionskip}{-0.5cm}  
	\includegraphics[width=3.5in]{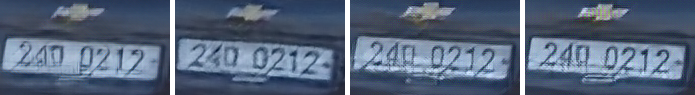}
	\includegraphics[width=3.5in]{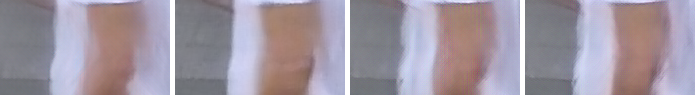}
	\includegraphics[width=3.5in]{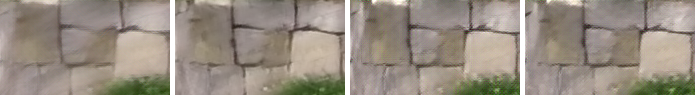}
	\caption{From left to right, we show the results of model \textbf{B}, \textbf{BE}, \textbf{BC}, and \textbf{Proposed}, respectively.}
	\label{fig3}
\end{figure}
\vspace{-3mm}
\subsection{Comparisons with State-of-the-art Methods}
Since the problem of our model processing is dynamic deblurring caused by camera shake and object motion, this is different from the application scenario of the traditional uniform deblurring model. We evaluated the performance of our model in the GoPro dataset and K\"ohler benchmark\cite{Rolf2012Recording}, mainly compared to the kernel-based and GAN-based deblurring method.  

We first evaluated the model using the 1111 test image pairs provided by the GoPro dataset. Sun et al. used CNN to estimate blur kernels and used traditional deconvolution methods to recover sharp images. However, in most real-world scenarios, blur kernels are complex and difficult to predict, so they do not perform well on some nonlinear blurred images of the dataset. The model proposed by Nah et al. has good performance on the dataset, but there are still artifacts in some images. In addition, we also evaluated the performance of deblurGAN on this dataset. Deblured result from test on GoPro dataset are shown in Fig.\ref{fig2} and quantitative results are shown in the Table II. 
%由于我们的模型是处理的相机抖动与物体运动等造成的动态去模糊问题，这与传统的均匀去模糊模型的应用场景不相同。我们主要与基于深度学习的去模糊方法进行比较。We evaluate the performance of our model in the GOPRO dataset and Köhler Dataset.
%我们首先使用GOPRO数据集提供的1111对测试图像进行了性能评估。Sun等人使用CNN估计模糊核，并使用传统的反褶积方法来恢复清晰图像。但是在大多数真实场景下，模糊核是复杂且难以预估的，所以在该数据集一些非线性模糊图像上表现不佳。Nah等人提出了GOPRO数据集，他们的模型在该数据集上具有良好的表现，但是部分图像中仍然存在伪影。此外，我们还评估了deblurGAN在该数据集上的表现。
%Kohler dataset consists of 4 latent images and 12 differently blurred images for each of them. The blurs are caused by replaying recorded 6D camera motion, assuming linear CRF.This is a standard benchmark dataset for evaluation of blind deblurring algorithms.Results are in 表X

\begin{table}[htb]
	\centering
	\footnotesize 
	\caption{Quantitative results on test dataset}
	\renewcommand{\arraystretch}{1.3}

\begin{tabular}{c|p{15mm}|c|c|c|c|c}
	\hline
	\multicolumn{2}{c|}{\multirow{2}{*}{Methods}} & \multicolumn{2}{c|}{GoPro} & \multicolumn{2}{c|}{K\"ohler} & \multirow{2}{*}{Time} \\ \cline{3-6}
	\multicolumn{2}{c|}{} & PSNR & SSIM & PSNR & MSSIM &  \\ \hline
	\multirow{3}{*}{\begin{tabular}[c]{@{}c@{}}kernel\\ based\end{tabular}} & Xu\cite{6} & 25.18 & 0.896 & \textbf{27.47} & 0.811 & 2.87s \\
	& Pan\cite{9} & 26.76 & 0.849 & 26.47 & 0.794 & 18min \\
	& Sun\cite{11} & 24.64 & 0.843 & 25.22 & 0.774 & 20min \\ \hline
	\multirow{4}{*}{\begin{tabular}[c]{@{}c@{}}GAN\\ based\end{tabular}} & Nah\cite{16} & 29.08 & 0.914 & 26.48 & 0.808 & 2.87s \\
	& Ram.\cite{23} & 28.94 & 0.922 & 27.08 & 0.812 & - \\
	& Kupyn\cite{17} & 28.70 & 0.858 & 26.10 &\textbf{0.816}  &\textbf{0.59s}  \\ \cline{2-7} 
	& Ours & \textbf{29.32} & \textbf{0.933} & 26.55 & 0.811 & 0.64s \\ \hline
\end{tabular}

\end{table}
\vspace{-5mm}
\section{Conclusion}
In this letter, we introduce high frequency information of images and images of different scales to improve the quality of deblurred images. The proposed network uses the “coarse-to-fine” scheme to restore sharp images in an end-to-end manner which does not involve blur kernel estimation and therefore has a faster running speed. We also design a new content loss that takes into account both the overall quality and local details of the restored image. Experimental results show that the proposed network has excellent performance in deblurring tasks.
\bibliographystyle{IEEEtran}
\bibliography{reference}

\begin{thebibliography}{10}
\providecommand{\url}[1]{#1}
\csname url@samestyle\endcsname
\providecommand{\newblock}{\relax}
\providecommand{\bibinfo}[2]{#2}
\providecommand{\BIBentrySTDinterwordspacing}{\spaceskip=0pt\relax}
\providecommand{\BIBentryALTinterwordstretchfactor}{4}
\providecommand{\BIBentryALTinterwordspacing}{\spaceskip=\fontdimen2\font plus
\BIBentryALTinterwordstretchfactor\fontdimen3\font minus
  \fontdimen4\font\relax}
\providecommand{\BIBforeignlanguage}[2]{{%
\expandafter\ifx\csname l@#1\endcsname\relax
\typeout{** WARNING: IEEEtran.bst: No hyphenation pattern has been}%
\typeout{** loaded for the language `#1'. Using the pattern for}%
\typeout{** the default language instead.}%
\else
\language=\csname l@#1\endcsname
\fi
#2}}
\providecommand{\BIBdecl}{\relax}
\BIBdecl

\bibitem{1}
T.~F. Chan and C.-K. Wong, ``Total variation blind deconvolution,'' \emph{IEEE
  transactions on Image Processing}, vol.~7, no.~3, pp. 370--375, 1998.

\bibitem{2}
R.~Fergus, B.~Singh, A.~Hertzmann, S.~T. Roweis, and W.~T. Freeman, ``Removing
  camera shake from a single photograph,'' in \emph{ACM transactions on
  graphics}, vol.~25, no.~3.\hskip 1em plus 0.5em minus 0.4em\relax ACM, 2006,
  pp. 787--794.

\bibitem{3}
A.~Levin, Y.~Weiss, F.~Durand, and W.~T. Freeman, ``Understanding and
  evaluating blind deconvolution algorithms,'' in \emph{IEEE Conference on
  Computer Vision and Pattern Recognition}.\hskip 1em plus 0.5em minus
  0.4em\relax IEEE, 2009, pp. 1964--1971.

\bibitem{4}
------, ``Efficient marginal likelihood optimization in blind deconvolution,''
  in \emph{IEEE Conference on Computer Vision and Pattern Recognition}.\hskip
  1em plus 0.5em minus 0.4em\relax IEEE, 2011, pp. 2657--2664.

\bibitem{5}
D.~Krishnan, T.~Tay, and R.~Fergus, ``Blind deconvolution using a normalized
  sparsity measure,'' in \emph{IEEE Conference on Computer Vision and Pattern
  Recognition}.\hskip 1em plus 0.5em minus 0.4em\relax IEEE, 2011, pp.
  233--240.

\bibitem{6}
L.~Xu, S.~Zheng, and J.~Jia, ``Unnatural l0 sparse representation for natural
  image deblurring,'' in \emph{IEEE conference on computer vision and pattern
  recognition}, 2013, pp. 1107--1114.

\bibitem{9}
J.~Pan, D.~Sun, H.~Pfister, and M.-H. Yang, ``Blind image deblurring using dark
  channel prior,'' in \emph{IEEE Conference on Computer Vision and Pattern
  Recognition}, 2016, pp. 1628--1636.

\bibitem{10}
S.~A. Bigdeli, M.~Zwicker, P.~Favaro, and M.~Jin, ``Deep mean-shift priors for
  image restoration,'' in \emph{Advances in Neural Information Processing
  Systems}, 2017, pp. 763--772.

\bibitem{11}
J.~Sun, W.~Cao, Z.~Xu, and J.~Ponce, ``Learning a convolutional neural network
  for non-uniform motion blur removal,'' in \emph{IEEE Conference on Computer
  Vision and Pattern Recognition}, 2015, pp. 769--777.

\bibitem{12}
C.~J. Schuler, M.~Hirsch, S.~Harmeling, and B.~Sch{\"o}lkopf, ``Learning to
  deblur,'' \emph{IEEE transactions on pattern analysis and machine
  intelligence}, vol.~38, no.~7, pp. 1439--1451, 2015.

\bibitem{13}
L.~Li, J.~Pan, W.-S. Lai, C.~Gao, N.~Sang, and M.-H. Yang, ``Learning a
  discriminative prior for blind image deblurring,'' in \emph{IEEE Conference
  on Computer Vision and Pattern Recognition}, 2018, pp. 6616--6625.

\bibitem{14}
L.~Xu, J.~S. Ren, C.~Liu, and J.~Jia, ``Deep convolutional neural network for
  image deconvolution,'' in \emph{Advances in Neural Information Processing
  Systems}, 2014, pp. 1790--1798.

\bibitem{15}
F.~Couzinie-Devy, J.~Sun, K.~Alahari, and J.~Ponce, ``Learning to estimate and
  remove non-uniform image blur,'' in \emph{IEEE Conference on Computer Vision
  and Pattern Recognition}, 2013, pp. 1075--1082.

\bibitem{2014Generative}
I.~J. Goodfellow, J.~Pouget-Abadie, M.~Mirza, B.~Xu, D.~Warde-Farley, S.~Ozair,
  A.~Courville, and Y.~Bengio, ``Generative adversarial networks,''
  \emph{Advances in Neural Information Processing Systems}, vol.~3, pp.
  2672--2680, 2014.

\bibitem{18}
C.~Ledig, L.~Theis, F.~Husz{\'a}r, J.~Caballero, A.~Cunningham, A.~Acosta,
  A.~Aitken, A.~Tejani, J.~Totz, Z.~Wang \emph{et~al.}, ``Photo-realistic
  single image super-resolution using a generative adversarial network,'' in
  \emph{IEEE Conference on Computer Vision and Pattern Recognition}, 2017, pp.
  4681--4690.

\bibitem{19}
S.~Iizuka, E.~Simo-Serra, and H.~Ishikawa, ``Globally and locally consistent
  image completion,'' \emph{ACM Transactions on Graphics}, vol.~36, no.~4, p.
  107, 2017.

\bibitem{20}
J.-Y. Zhu, T.~Park, P.~Isola, and A.~A. Efros, ``Unpaired image-to-image
  translation using cycle-consistent adversarial networks,'' in \emph{IEEE
  Conference on Computer Vision and Pattern Recognition}, 2017, pp. 2223--2232.

\bibitem{16}
S.~Nah, T.~Hyun~Kim, and K.~Mu~Lee, ``Deep multi-scale convolutional neural
  network for dynamic scene deblurring,'' in \emph{IEEE Conference on Computer
  Vision and Pattern Recognition}, 2017, pp. 3883--3891.

\bibitem{17}
O.~Kupyn, V.~Budzan, M.~Mykhailych, D.~Mishkin, and J.~Matas, ``Deblurgan:
  Blind motion deblurring using conditional adversarial networks,'' in
  \emph{IEEE Conference on Computer Vision and Pattern Recognition}, 2018, pp.
  8183--8192.

\bibitem{23}
S.~Ramakrishnan, S.~Pachori, A.~Gangopadhyay, and S.~Raman, ``Deep generative
  filter for motion deblurring,'' in \emph{IEEE International Conference on
  Computer Vision}, 2017, pp. 2993--3000.

\bibitem{21}
P.~Isola, J.-Y. Zhu, T.~Zhou, and A.~A. Efros, ``Image-to-image translation
  with conditional adversarial networks,'' in \emph{IEEE Conference on Computer
  Vision and Pattern Recognition}, 2017, pp. 1125--1134.

\bibitem{22}
G.~Huang, Z.~Liu, L.~Van Der~Maaten, and K.~Q. Weinberger, ``Densely connected
  convolutional networks,'' in \emph{IEEE Conference on Computer Vision and
  Pattern Recognition}, 2017, pp. 4700--4708.

\bibitem{2016Perceptual}
J.~Johnson, A.~Alahi, and L.~Fei-Fei, ``Perceptual losses for real-time style
  transfer and super-resolution,'' in \emph{European conference on computer
  vision}.\hskip 1em plus 0.5em minus 0.4em\relax Springer, 2016, pp. 694--711.

\bibitem{2016Deep}
S.~Su, M.~Delbracio, J.~Wang, G.~Sapiro, W.~Heidrich, and O.~Wang, ``Deep video
  deblurring for hand-held cameras,'' in \emph{IEEE Conference on Computer
  Vision and Pattern Recognition}, 2017, pp. 1279--1288.

\bibitem{2017Deep}
N.~Xu, B.~Price, S.~Cohen, and T.~Huang, ``Deep image matting,'' in \emph{IEEE
  Conference on Computer Vision and Pattern Recognition}, 2017, pp. 2970--2979.

\bibitem{2014Coarse}
J.~Zhang, S.~Shan, M.~Kan, and X.~Chen, ``Coarse-to-fine auto-encoder networks
  (cfan) for real-time face alignment,'' in \emph{European Conference on
  Computer Vision}.\hskip 1em plus 0.5em minus 0.4em\relax Springer, 2014, pp.
  1--16.

\bibitem{he2016deep}
K.~He, X.~Zhang, S.~Ren, and J.~Sun, ``Deep residual learning for image
  recognition,'' in \emph{Proceedings of the IEEE conference on computer vision
  and pattern recognition}, 2016, pp. 770--778.

\bibitem{2016Instance}
D.~Ulyanov, A.~Vedaldi, and V.~Lempitsky, ``Instance normalization: The missing
  ingredient for fast stylization,'' \emph{arXiv preprint arXiv:1607.08022},
  2016.

\bibitem{2015U}
O.~Ronneberger, P.~Fischer, and T.~Brox, ``U-net: Convolutional networks for
  biomedical image segmentation,'' in \emph{International Conference on Medical
  image computing and computer-assisted intervention}.\hskip 1em plus 0.5em
  minus 0.4em\relax Springer, 2015, pp. 234--241.

\bibitem{2017High}
T.-C. Wang, M.-Y. Liu, J.-Y. Zhu, A.~Tao, J.~Kautz, and B.~Catanzaro,
  ``High-resolution image synthesis and semantic manipulation with conditional
  gans,'' in \emph{IEEE Conference on Computer Vision and Pattern Recognition},
  2018, pp. 8798--8807.

\bibitem{shi2016real}
W.~Shi, J.~Caballero, F.~Husz{\'a}r, J.~Totz, A.~P. Aitken, R.~Bishop,
  D.~Rueckert, and Z.~Wang, ``Real-time single image and video super-resolution
  using an efficient sub-pixel convolutional neural network,'' in \emph{IEEE
  conference on Computer Vision and Pattern Recognition}, 2016, pp. 1874--1883.

\bibitem{2017Wasserstein}
M.~Arjovsky, S.~Chintala, and L.~Bottou, ``Wasserstein gan,'' \emph{arXiv
  preprint arXiv:1701.07875}, 2017.

\bibitem{2017Improved}
I.~Gulrajani, F.~Ahmed, M.~Arjovsky, V.~Dumoulin, and A.~C. Courville,
  ``Improved training of wasserstein gans,'' in \emph{Advances in Neural
  Information Processing Systems}, 2017, pp. 5767--5777.

\bibitem{Rolf2012Recording}
R.~K{\"o}hler, M.~Hirsch, B.~Mohler, B.~Sch{\"o}lkopf, and S.~Harmeling,
  ``Recording and playback of camera shake: Benchmarking blind deconvolution
  with a real-world database,'' in \emph{European Conference on Computer
  Vision}.\hskip 1em plus 0.5em minus 0.4em\relax Springer, 2012, pp. 27--40.

\end{thebibliography}
\end{document}